\documentclass[twocolumn,fleqn,natbib]{svjour2}
\bibpunct{[}{]}{;}{n}{}{,} 
\smartqed  
\usepackage{graphicx}
\usepackage{amsmath}
\usepackage{color}

\journalname{Granular Matter}
\begin{document}

\title{Effect of the granular material on the maximum holding force of a granular gripper}

\titlerunning{Effect of material on a granular gripper}        

\author{Juli\'an M. G\'omez--Paccapelo \and Angel A. Santarossa \and H. Daniel Bustos \and Luis A. Pugnaloni
} \institute{Juli\'an M. G\'omez-Paccapelo \and Angel A. Santarossa \and H. Daniel Bustos: Dpto. F\'isica, FCEyN, Universidad Nacional de La Pampa, Uruguay 151, 6300 Santa Rosa, La Pampa, Argentina.
               \\
             L. A. Pugnaloni: Dpto. F\'isica, FCEyN, Universidad Nacional de La Pampa, Uruguay 151, 6300 Santa Rosa, La Pampa, CONICET, Argentina. E-mail: luis.pugnaloni@exactas.unlpam.edu.ar \\
             Angel A. Santarossa (current address): Institute for Multiscale Simulation, Universit\"at Erlangen-N\"urnberg, Cauerstra\ss{}e 3, 91058 Erlangen, Germany.  
}

\authorrunning{J. M. G\'omez--Paccapelo et al.} 

\date{Received: date }

\maketitle

\begin{abstract}
 A granular gripper is a device used to hold objects by taking advantage of the phenomenon of Reynold's dilatancy. A membrane containing a granular sample is allowed to deform around the object to be held and then vacuum is used to jam the granular material inside the membrane. This allows to hold the object against external forces since deformation of the granular material is prevented by not allowing the system to increase its volume. The maximum holding force supported by the gripper depends on a number of variables. In this work, we show that in the regime of frictional holding (where the gripper does not interlock with the object), the maximum holding force does not depend on the granular material used to fill the membrane. Results for a variety of granular materials can be collapsed into a single curve if maximum holding force is plotted against the penetration depth achieved. The results suggest that the most important feature in selecting a particular granular material is its deformability to ensure an easy flow during the initial phase of the gripping process.    
\end{abstract}

\section{Introduction}\label{sec:into} 

The handling of objects is a regular task in the industry. Holding objects of well defined size, shape and hardness can be done by robotic arms that present a gripper with a matching shape that fits the object to hold (e.g., hooks). Also, magnets can be used with ferromagnetic objects, and suction systems with objects presenting smooth surfaces. However, matching some of the gripping characteristics of the human hand is always desirable. Fingered grippers have been developed systematically over decades to provide a more universal (any shape, size and hardness) gripping ability (for a recent review see Ref. \cite{Shintake2018}). However, these hand-like grippers require complex auxiliary systems to asses the gripping problem and take multiple decisions on how to handle each finger (time and amplitude of aperture, time of closure, applied pressure to hold, etc.). These auxiliary systems require complex hardware and software.  

More than 30 years ago, there where new proposals to tackle the gripping problem by using some unique properties of dens granular matter \cite{schmidt1978,perovskii1980,rienmuller1988}. The ``granular gripper'' consists in a flexible impermeable bag partially filled with a granular material and connected to a vacuum pump. When the interior is at atmospheric pressure, the bag (and its material inside) can be easily deformed and reshaped. Simply pressing the bag against an object makes the bag to deform, partially conforming to the object shape. When vacuum is applied inside the bag, this contracts and confines the granular sample, which becomes rigid. If an object had been partially wrapped by the bag, the new solid state of the bag will cause the object to be gripped. There are three properties of the granular material inside the bag that allow this technology to work. First, the flowability of the grains when there is no vacuum applied \cite{Amend2016}, then the jamming of the granular sample which sustains the external pressure when vacuum is applied \cite{Jaeger2014}, and finally the Reynold's dilatancy \cite{Reynolds1886,Sakaie2008} which causes the sample to become hard to deform since the external pressure prevents the volume increase needed by the grains to pass each other during shear.   

In recent years, the interest in granular grippers has increased significantly, partially due to the study about the gripping mechanisms and their connection to the mechanical strength of the gripper by Brown et al. \cite{brown2010}. In that paper, the authors describe three gripping mechanism: (i) friction, (ii) suction, and (iii) interlocking. Interlocking requires the gripper to wrap the object (or some protrusion of it) to the extent that detaching the object would require a large deformation of part of the rigidized bag. Interlocking is the most effective gripping mechanism, but is not achieved in most objects by simply pressing the fluid bag against the object. Therefore, this mechanism is far from being universal in practice. Suction is caused by the formation of a sealed cavity between the object and the gripper while the object is being pulled apart. This mechanism works only if sealing can readily occur, which is not possible with many rough surfaces. Friction is in fact the only universal mechanism at work under all gripping conditions. It is less effective than the other two mechanism, but more than sufficient to sustain the weight of objects of moderate size and density \cite{brown2010}.       

Since the extent of wrapping is essential to achieve a high maximum holding force, some simple techniques can be used to aid this process. One such technique is to partially inflate the bag before approaching the object \cite{amend2012}. This gives more available volume to the granular sample, which eases the flow around the object. This so-called ``positive pressure gripper'' can conform to an object applying up to 90\% less force on it. In the same spirit, Nishida et al. showed that the maximum holding force increases if some extra space is left in the bag for the material to flow during the conforming phase \cite{nishida2014}. 

The most straightforward method to enhance wrapping is pressing the gripper against the target object with higher forces. This applied force $F_{\rm a}$ is called the activation force. Brown et al. \cite{brown2010} considered the maximum holding force $F_{\rm h}$ (the maximum force that the gripper can support before the object is detached when pulled axially) as a function of the maximum contact angle between the object and the bag. The contact angle is a suitable measure of the extent of wrapping. However, in industrial applications, the angle of contact is difficult to measure. A more natural choice is to furnish the gripper with a force sensor to measure $F_{\rm a}$. Therefore, for practical applications one would require to know the $F_{\rm h}-F_{\rm a}$ curve of a gripper to be able to predict the necessary activation force to hold a given weight. The  $F_{\rm h}-F_{\rm a}$ curve will depend on constructive details of the gripper, the granular material used and the size and shape of the target object.  

Brown et al. mentioned that the details of the granular material seem to have a minor role in the maximum holding force attained by the gripper as long as the material used does not interfere with the gripper conforming the object \cite{brown2010}. However, more recent studies seem to  show that the holding force depends on the granular material used inside the bag \cite{nishida2014,meuleman2017}. In this work, we revisit this issue by performing a series of experiments with different granular materials. We show that for grain sizes below $1/15$ of the target object diameter, the actual material used has an important effect on the $F_{\rm h}-F_{\rm a}$ curve. However, we find that this is only due to the flowability of the material while conforming the target object as claimed by Brown et al. \cite{brown2010}. When $F_{\rm h}$ is plotted against the penetration depth of the object inside the gripper bag all data collapse into a single curve. We also show that for larger grains this collapse fails; and we discuss plausible explanations for such deviation from the collapsing data.

\begin{figure}[t]
	\centering
	\includegraphics[width=0.8\linewidth]{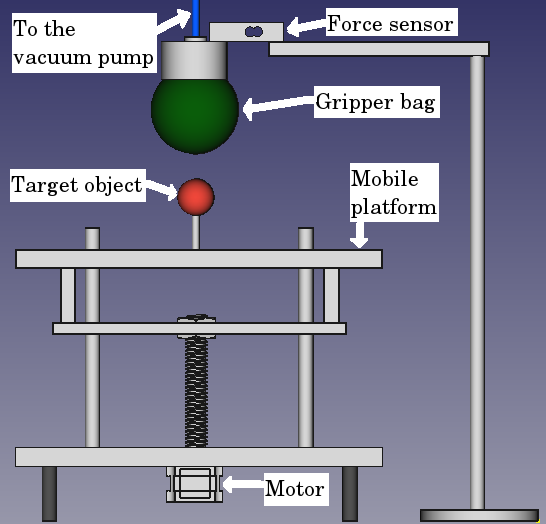}
	\caption{Sketch of the experimental apparatus.}
	\label{fig:exp}
\end{figure}

\section{Experimental setup}\label{sec:NumSetup}

Figure \ref{fig:exp} shows a sketch of the apparatus. The granular gripper is composed of a rubber bag partially filled with a granular sample. The bag is attached to a Teflon fitting that connects the interior of the bag to a vacuum pump (Leybold Vp2, $90$ kPa maximum differential pressure) through a flexible pipe. The Teflon part is connected to a force sensor (CT460B, maximum force $\pm 50$ N), which is fixed to a rigid I-shape beam.

Beneath the gripper, the object to be gripped (a glass sphere of $17$ mm in diameter coated in rubber) is attached to a movable base. The base can be displaced up and down at constant velocity ($2.8 \pm 0.4$ mm/s) by means of a screw connected to a motor (BOSCH FPG). The motion of the base is controlled by means of an Arduino microcontroller using the feedback from the force sensor that sustains the gripper. The motion of the platform emulates the action of a robotic arm, where the object is handled in a controlled way instead of the griper, which remains fixed. 

We define the activation force as the maximum force exerted in the vertical direction by the object onto the gripper while deforming the bag around it. The upward motion of the object can be stopped for any prescribed value of the activation force. The maximum holding force is defined as the critical force at which the object is detached from the gripper during the downward motion.

The protocol for any single measurement is as follows. (i) The bag is inflated for a few seconds using a small positive pressure to allow the granular material to relax and loose memory of previous manipulation of the bag. (ii) The bag inner pressure is let to equilibrate with ambient pressure. (iii) The object is elevated at constant velocity and pressed against the gripper bag until the vertical force between the object and the gripper reaches the desired activation force. (iv) The movable base is held static during $10$ s for the granular material to relax. (v) Negative pressure ($86 \pm 7$ kPa) is applied to the gripper so that the granular material becomes rigid inside and the object gets gripped. This pressure is held for $36$ s while the material relaxes. (vi) With the vacuum pressure fixed, the object is moved downward by the base until it is detached from the gripper. We measure the depth that the object has dipped into the bag in each case using a digital image of the configuration right after applying the vacuum pressure [i.e., after step (v)].

\begin{figure}[t]
\begin{flushright}
 \includegraphics[width=0.8\linewidth]{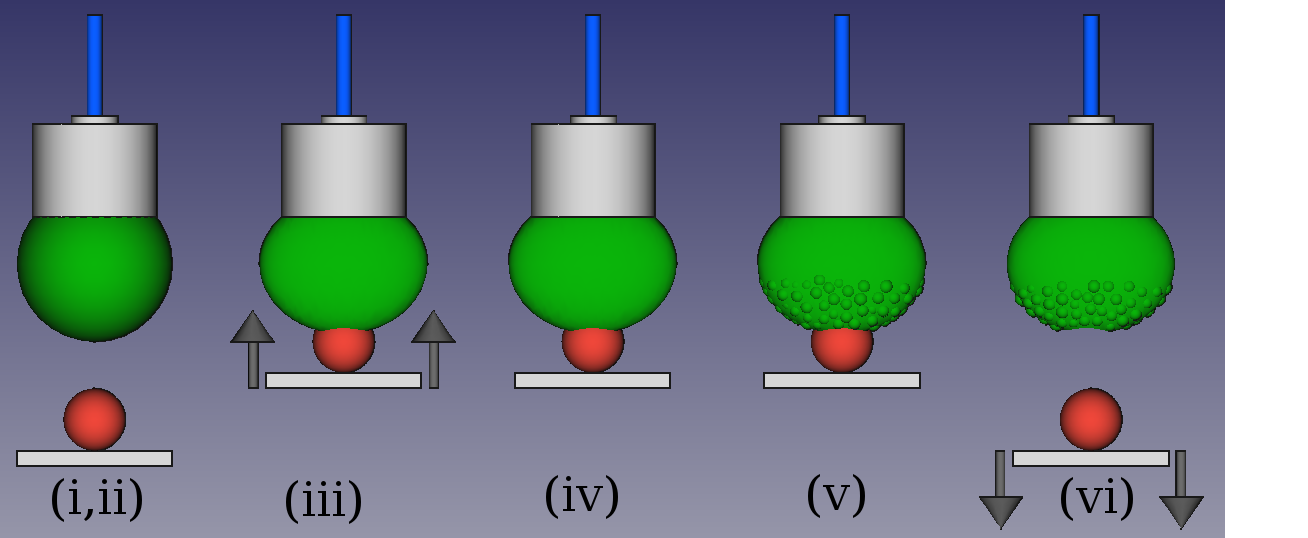}
	\includegraphics[width=1\linewidth]{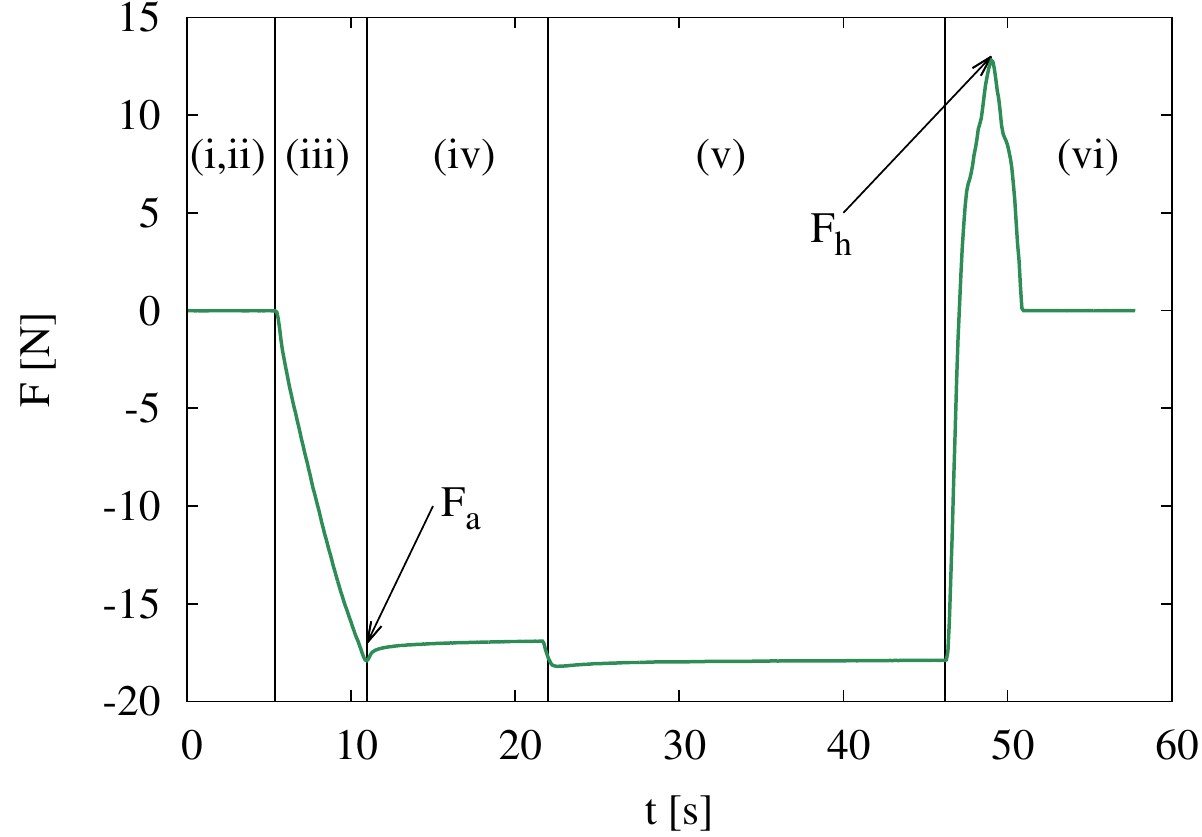}
\end{flushright}
	\caption{Force exerted by the gripper onto the target object during one experiment. The phases of the protocol are indicated in the figure: (i,ii) the bag is inflated for a few seconds and then the pressure is let to equilibrate with ambient pressure, (iii) the object is pressed against the gripper until the desired activation force, (iv) the object is held at its position for 10 s, (v) negative pressure is applied for 36 s, (vi) the object is pulled downward and detached from the gripper at constant speed. $F_{\rm h}$ indicates the maximum holding force and $F_{\rm a}$ is the activation force.}
	\label{fig:protocol}
\end{figure}

\begin{figure}[h]
	\centering
	\includegraphics[width=\linewidth]{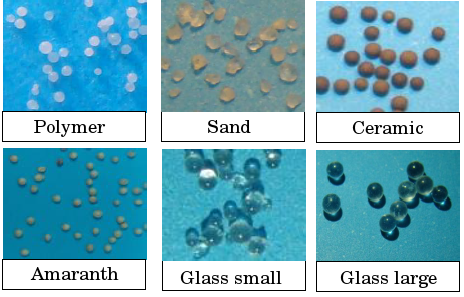}
	\caption{Samples of the granular materials used in the experiments: polymer microspheres, sand, ceramic beads, amaranth seeds, small glass microspheres and large glass beads.}
	\label{fig:materials}
\end{figure}

\begin{table}[h]
\centering
 \begin{tabular}{l r  r }
  \hline\hline
  Material & Density [kg/m$^3$] & Grain size [$\mu$m] \\
  \hline
  polymer & $940 \pm \,\ 40$ & $125-\,\ 212$  \\
  sand & $2590 \pm 110$ & $212-\,\ 600$ \\
  ceramic & $3160 \pm 130$ & $425-\,\ 850$ \\ 
  amaranth & $1340 \pm \,\ 90$& $1000-1500$ \\ 
  glass small & $2500 \pm 100$ & $200-\,\ 400$\\
  glass large & $2500 \pm 100$ & $2000-2500$ \\
  \hline\hline
 \end{tabular}
\caption{List of material properties of the granular materials used for the experiments.} \label{tab:materials}
\end{table}

During the entire experimental run, the force in the force sensor is registered at 100 samples/s with a resolution of $0.05$ N. In Fig. \ref{fig:protocol} we show an example of the force exerted on the target object during one experiment. Each phase in the protocol described above is indicated in the figure. During phase (iii) we observe the increase of the force on the object in the downward direction (negative forces) until the prescribed value for $F_{\rm a}$ is achieved. In the relaxation phase (iv) we observe a small decrease and rapid saturation of the absolute value of the force. In phase (v) the applied vacuum induce an small increase of the volume of the granular material inside the bag (Reynold's dilatancy), which leads to an slight increase in the absolute value of the force on the target object. When the object is pulled back down by the platform in phase (vi) the force on the object rapidly decreases in absolute value and becomes positive until the object detaches from the gripper and the force relaxes to zero. The maximum holding force $F_{\rm h}$ is extracted from the maximum in Fig. \ref{fig:protocol}. For any given activation force, we carried out between 5 and 10 realizations of the experiment. The standard deviation of the maximum holding force is usually below 5\% of the mean in all our experiments. 

We have tested different granular materials (see Fig. \ref{fig:materials} and Table \ref{tab:materials}). Since material density and packing fraction varies, we used in all cases the same apparent volume (60 cm$^3$) of material inside the gripper bag. The apparent volume was measured before pouring the material in the bag by filling a graduated tube with a funnel taking care of using always the same funnel position and filling speed. The materials chosen cover a wide range of particle sizes, material densities and stiffnesses. 

The target object is a glass sphere ($17.0$ mm in diameter) coated in the same rubbery material as the membrane used for the gripper bag. During each gripping experiment we take images with a CCD camera to measure the penetration depth $D$ of the target object into the gripper bag. This is measured after the vacuum has been applied and before pulling back the object. The relative vertical position of object and gripper is measured with $0.1$ mm resolution. In total, $100$ experiments were carried out including several realizations for different materials and different activation forces.

\section{Results}\label{sec:Results}

\begin{figure}[t]
	\centering
	\includegraphics[width=0.9\linewidth]{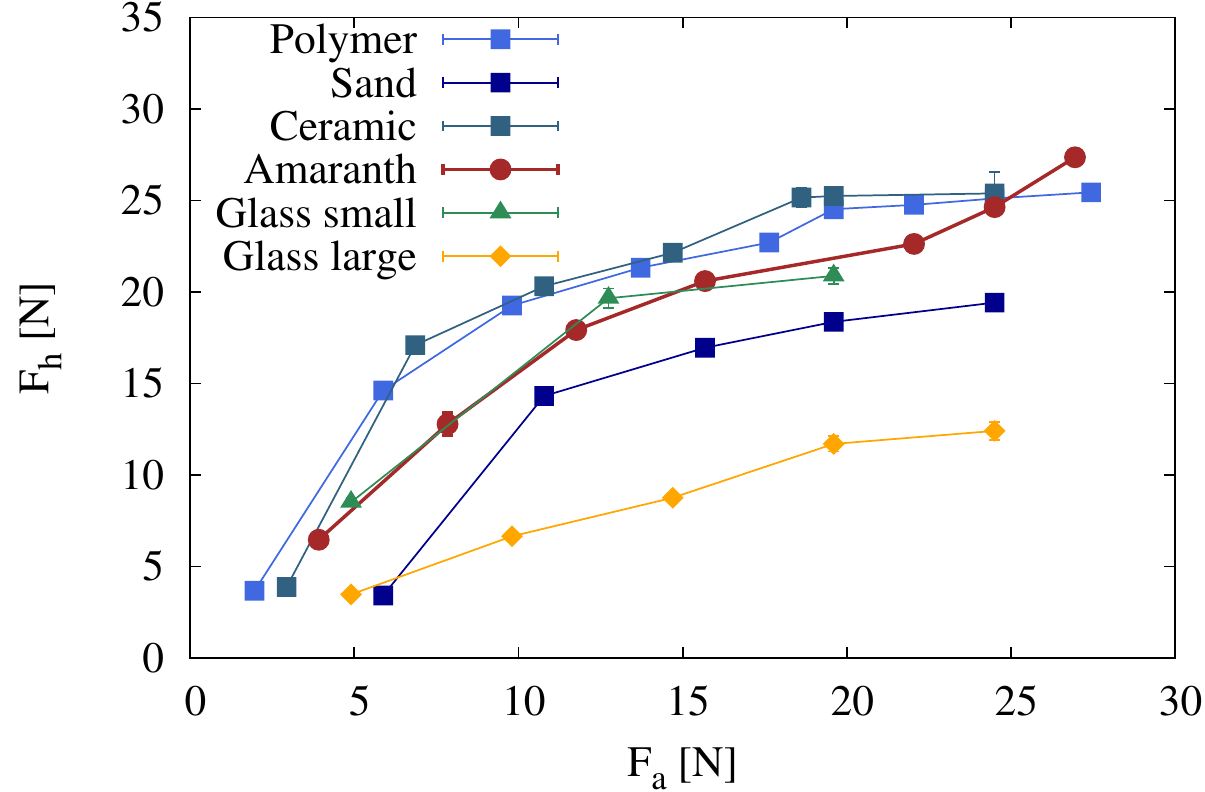}
	\caption{$F_{\rm h}$ as a function of $F_{\rm a}$ for various granular materials (see legend). Error bars correspond to the standard deviation.}
	\label{fig:fh-fa}
\end{figure}

We have carried out measurements of $F_{\rm h}$ for a range of $F_{\rm a}$. Figure \ref{fig:fh-fa} shows the results for all granular materials tested. Measurements are very reproducible with typical error in $F_{\rm h}$ below 5\%. The polymer microspheres, ceramic beads, small glass beads and amaranth seeds show similar results, although with some scatter. However, sand and large glass beads present a significantly lower $F_{\rm h}$ for any given $F_{\rm a}$. 

As we can see in Fig. \ref{fig:fh-fa},  $F_{\rm h}$ increases with $F_{\rm a}$ and saturates at around $30$ N. For some materials we where unable to reach the high values of  $F_{\rm a}$ required to achieve saturation due to limitations in the mechanical system. This saturation occurs because the gripper bag does not wrap the object completely covering it pass beneath the equator so that interlocking is at play. While increasing $F_{\rm a}$, the target object simply deepens into the bag creating a straight vertical cylindrical channel that does not close beneath the target object. The contact between the gripper and the object only occurs for the upper hemisphere of the object. Therefore, once the bag has covered the upper hemisphere, further penetration does not lead to any additional increase in the contact angle. It is worth mentioning that higher contact angles have been achieved in previous studies only by molding the gripper bag by hand \cite{brown2010}. It seems that proper interlocking cannot be attained without external intervention.  

\begin{figure}[t]	
	\centering
	\includegraphics[width=0.9\linewidth]{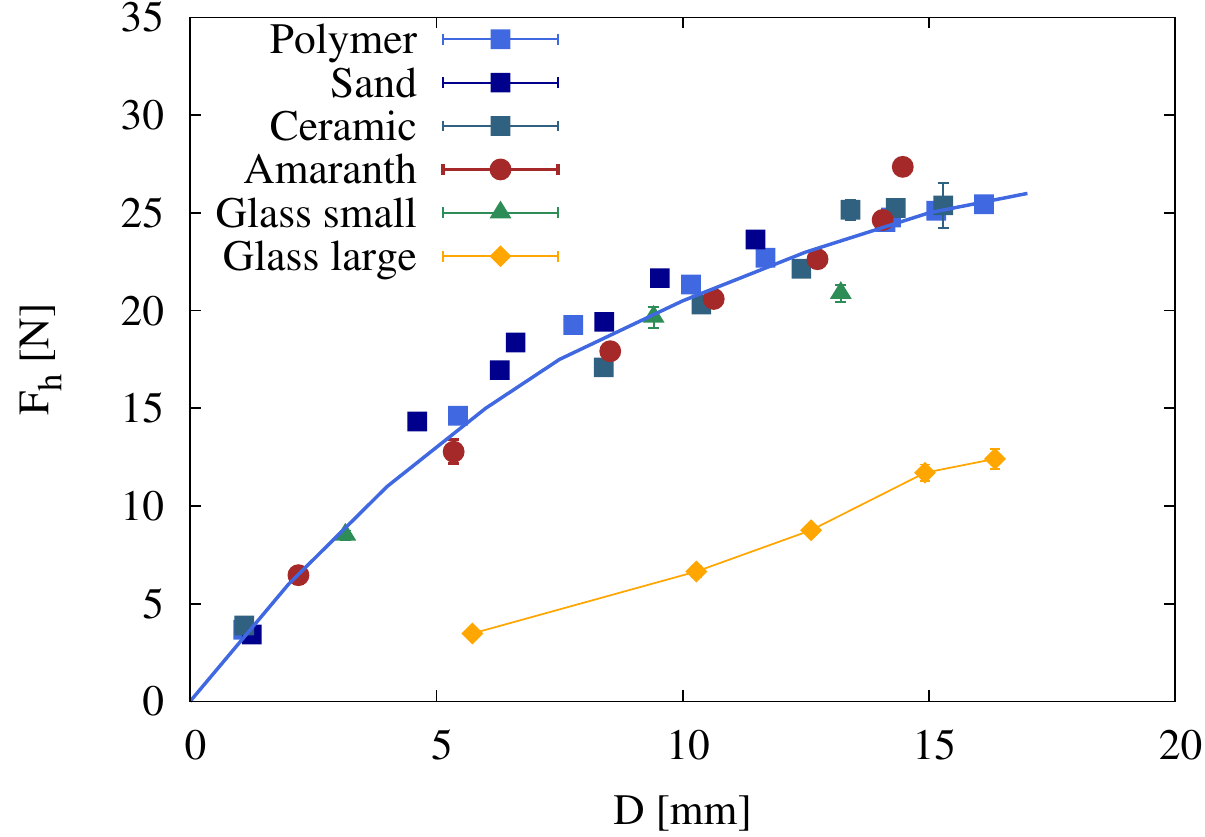}
	\caption{$F_{\rm h}$ as a function of penetration depth for various granular materials (see legend). Error bars correspond to the standard deviation.}
	\label{fig:fh-pd}
\end{figure}

Since it is impractical to measure the maximum contact angle, we used a different measure of the degree of wrapping. This is the penetration depth $D$ defined as the length that the object has penetrated into the bag for the given value of $F_{\rm a}$. In Fig. \ref{fig:fh-pd} we show $F_{\rm h}$ as a function of $D$. In this representation all data for small grain sizes collapse to a good degree for all materials tested. This result indicates that is in fact the penetration depth that controls the holding force. If a granular material displays a lower holding force for a given $F_{\rm a}$, this is simply caused by a less effective penetration.

\begin{figure}[b]
	\centering
	\includegraphics[width=0.8\linewidth]{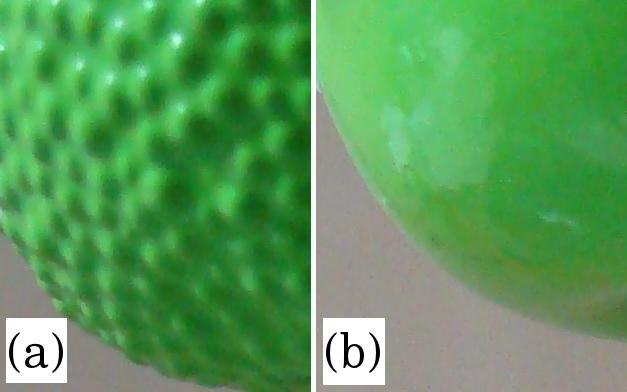}
	\caption{Photographs of the surface of the gripper bag while vacuum is applied for large glass beads (a) and for the polymer microspheres (b).}
	\label{fig:bumpy}
\end{figure}

The peculiar behavior observed for the large glass beads is worth of attention. In agreement with the lower $F_{\rm h}$ observed here for large grains, Amend et al. have reported that smaller mesh sizes for the granular material do lead to lower object retention \cite{Amend2016}. We have tested if materials with small grains present an extra contribution to the holding force due to suction that is not present for large grains. We did this by using a perforated target object that prevents the formation of a seal between object and gripper. However, suction seems to be negligible in our system.  Interestingly, we have observed that the bag, when vacuum is applied, presents a bumpy surface since it copies the shape of the granular sample inside. For large grains, this is particularly apparent (see Fig. \ref{fig:bumpy}). This makes the bag to contact the target object only at the protruding spots since the concave regions of the bag surface are deeper for large grains. Since the rubbery bags present some degree of adhesiveness, the holding force depends partly on the effective area of contact. As a consequence, large grain sizes induce a marked drop in holding force. This is consistent with recent experiments based on pin array grippers that show a clear dependence of $F_{\rm h}$ on the number of contact points between the gripper and the object \cite{Mo2019}.

\section{Discussion and conclusions}\label{Sec:Concl}

We have shown that the $F_{\rm h}-F_{\rm a}$ curve for a granular gripper is sensitive to the granular material used. From a practical perspective, this implies that a robotic arm that senses the activation force while conforming the gripper to the object will require a calibration curve for the particular granular material used. However, if the same penetration of the object is achieved for two different materials, then $F_{\rm h}$ becomes material independent. 

The previous observations are consistent with reports in the literature that seem at first sight contradictory. On the one hand, some workers found that the holding force is material dependent \cite{nishida2014,meuleman2017}. On the other hand, Brown et al. suggested that the granular material should play a marginal role on the holding force \cite{brown2010}. However, these studies where considering as a control variable either $F_{\rm a}$ \cite{nishida2014,meuleman2017} or the contact angle \cite{brown2010}. The contact angle is in fact a function of the penetration depth for the frictional mechanism of gripping studied here. We have shown that both claims are compatible because, even if $F_{\rm h}$ does not depend on the contact angle (or penetration depth)  as shown in Fig. \ref{fig:fh-pd}, the contact angle does depend on material properties for a given $F_{\rm a}$ due to the different flowability of the materials. 

We have observed that the collapse of the data fails for large grains; larger than  $1/15$ of the target object diameter. This seems to be connected to the fact that large grains create a bumpy surface on the gripper bag that reduces the effective contact area between the bag and the target object. This, in turn, leads to a marked drop in the maximum holding force. 

These findings suggest that a robotic arm capable of sensing the penetration depth can in fact use the master curve in Fig. \ref{fig:fh-pd} to estimate the maximum holding force at each gripping operation. This can be achieved by equipping the arm with a force sensor to detect the first contact with the target object and a displacement sensor to measure penetration from that position on. It is important to mention that the master curve shown in Fig. \ref{fig:fh-pd} needs to be obtained for each object size and shape. The universal character of this curve is with respect to the granular material only.

\textbf{Acknowledgments:} We thank N. Arce and J. P. Cagnola from Universidad Tecnol\'ogica Nacional (La Plata) for their contribution in the design and test of the experimental apparatus. We are indebted to G. Corral and M. Baccin for their help during the experiments. This work has been supported in part by ANPCyT (Argentina) through grant PICT-2016-2658, UTN (Argentina) through grant PID-MAUTNLP-4415 and FCEyN-UNLPam through grant F-55.

\textbf{Contributions:} JMGP and AAS have contributed equally to this work.

\textbf{Conflict of Interest:} The authors declare that they have no conflict of interest.

\end{document}